\documentclass[11pt]{article}
\usepackage[letter]{preamble}

\begin{document}

\title{
  \bfseries Optimal Preprocessing for Answering \\ On-Line Product Queries\thanks{This manuscript appeared originally as	TR 71/87, the Moise and Frida Eskenasy Institute of Computer Science, Tel Aviv University (1987).}}

\author{
  Noga Alon\thanks{The research of this author was supported
  in part by Allon Fellowship, Bat Sheva de Rothschild Foundation
  and the Fund for Basic Research administered by the Israel Academy
  of Sciences.}}
\affil{
  \vspace*{-8pt}
  Department of Mathematics\\
  School of Mathematical Sciences\\
  Tel Aviv University\\
  Tel Aviv, Israel 69978}
\author{
  Baruch Schieber\thanks{The research of this author was supported
  by the Applied Mathematical Sciences
  subprogram of the Office of Energy Research, U.S. Department of
  Energy under contract number DE-AC02-76ER03077.}}
\affil{
  \vspace*{-8pt}
  Department of Computer Science\\
  School of Mathematical Sciences\\
  Tel Aviv University\\
  Tel Aviv, Israel 69978}

\date{}
\maketitle

\begin{abstract}
We examine the amount of preprocessing needed for answering
certain
on-line queries as fast as possible. We start with the following
basic problem. Suppose we are given a semigroup $(S,\myotimes )$.
Let $s_1 ,\ldots, s_n$ be elements of $S$. We want to answer on-line
queries of the form,
``What is the product $s_i \myotimes s_{i+1} \myotimes \cdots \myotimes s_{j-1} \myotimes s_j$?''
for any given $1\le i\le j\le n$.
We show that a preprocessing of
$\bigtheta{n \lambda (k,n)}$ time and space
is both necessary and sufficient to answer
each such query in at most $k$ steps, for any fixed $k$.
The function $\lambda (k,\cdot)$ is the inverse
of a certain function at the $\floor{k/2}$-th
level of the primitive recursive
hierarchy. In case
linear preprocessing is desired, we show that one can
answer each
such query in $\bigo{ \alpha (n)}$ steps and that this is best
possible. The function $\alpha (n)$ is the
inverse Ackermann function.

We also consider the following extended problem.
Let $T$ be a tree with an element of $S$ associated with
each of its vertices. We want to answer on-line
queries of the form,
``What is the product of the elements associated with the vertices
along the path from $u$ to $v$?''
for any pair of vertices $u$ and $v$ in $T$.
We derive results that are similar to the above, for the
preprocessing needed for answering such queries.

All our sequential preprocessing
algorithms can be parallelized efficiently to give optimal
parallel algorithms which run in $\bigo{\log n}$ time on a CREW PRAM.
These parallel algorithms are optimal in both running time
and total number of operations.

Our algorithms, especially for the semigroup of the real numbers
with the minimum or maximum operations, have various applications
in certain graph algorithms, in the utilization of communication
networks and in Database retrieval.
\end{abstract}

\section{Introduction}
We examine the amount of preprocessing needed for answering certain
on-line queries as fast as possible.
Suppose we are given a semigroup $(S,\myotimes )$.
We consider the following queries.

\par\noindent\textit{Linear Product Query.}
Let $s_1 ,\dots, s_n$ be elements of $S$. We want to answer on-line
queries of the form, ``What is the product
$s_i \myotimes s_{i+1} \myotimes \cdots \myotimes s_{j-1} \myotimes s_j$?'' for any given
$1\le i\le j\le n$.

\par\noindent\emph{Tree Product Query.}
Let $T$ be an unrooted tree with an element of $S$ associated with
each of its vertices. We want to answer on-line
queries of the form, ''What is the product
of the elements associated with the vertices along the path from
$u$ to $v$?'' for any pair of vertices $u$ and $v$ in $T$.

We present very efficient preprocessing
algorithms for the above queries and show that under reasonable
assumptions these algorithms are best possible. Our assumptions
are as follows.
(1) Given any two elements $a$ and $b$ in the semigroup $S$
we can compute $a \myotimes b$ in constant time.
(2) The only available operation on the semigroup elements
is the semigroup operation.

\noindent
We say that we answer a query in $k$ \emph{steps} if we
have to multiply $k$ available precomputed
semigroup elements to get the answer.

We achieve the following results.

\par\noindent\textit{Linear Product Query.}
We show that in order to answer each Linear Product query in at most
$k$ steps, for any fixed $k$, a preprocessing of
$\bigtheta{n \lambda (k,n)}$ time and space is both necessary and
sufficient. The function $\lambda (k,\cdot)$ is the
inverse of a certain function at the $\floor{k/2}$-th
level of the primitive recursive hierarchy.
We also present a linear preprocessing algorithm
that enables us to answer each query in
$\bigo{\alpha (n)}$ steps, where $\alpha (n)$ is the inverse  Ackermann
function (cf.\cite{Ac-28}).
It is further shown that no linear preprocessing algorithm can do
better.

\par\noindent\emph{Tree Product Query.}
We show that in order to answer each Tree Product query in at most
$2k$ steps, for any fixed $k$, a preprocessing of
$\bigo{n \lambda (k,n)}$ time and space is sufficient, where $n$ is the
number of vertices. (From the
previous result we have that a preprocessing of
$\bigomega{n \lambda (2k,n)}$ time and space is necessary for answering
Tree Product queries in certain trees.) Here also
the best linear preprocessing algorithm
enables us to answer each query in
$\bigtheta{\alpha (n)}$ steps.

All our sequential preprocessing algorithms can be parallelized
efficiently. The models of parallel computation
are the well-known concurrent-read exclusive-write (CREW)
parallel random access machine (PRAM) and
concurrent-read concurrent-write (CRCW) PRAM. (See, e.g.,~\cite{Vi-83}.)
The resulting parallel preprocessing algorithms
for answering Linear Product queries
in at most $k$ steps and for answering
Tree Product queries in at most $2k$ steps
run in $\bigo{\log n}$ time using $n \lambda (k,n) / \log n$
processors on a CREW PRAM.
These algorithms are the fastest algorithms
achievable on a CREW PRAM. This is, since $\bigomega{\log n}$
is a lower bound on the running time of any parallel preprocessing
algorithm for answering product queries in $\smallo{\log n}$
steps on a CREW PRAM. This lower bound is easily derived from the
lower bound of~\cite{CD-82}.
For the case where the semigroup operation is the
maximum operation (or similar) we can achieve a parallel preprocessing
algorithm for the Linear Product problem that runs in
$\bigo{\log\log n}$ time using $n \lambda (k,n) / \log \log n$
processors on a CRCW PRAM.
This algorithm is the fastest algorithm
achievable with a linear number of processors on a CRCW PRAM, as
proved in~\cite{Va-75}.
Notice that all these parallel algorithms are also optimal in
their total number of operations (i.e., the product
of their running time and number of processors). This is since
their total number of operations is equal to the
time complexity of the corresponding sequential algorithms.

The parallel preprocessing algorithms use the same ideas as the
sequential algorithms together with the
parallel preprocessing algorithm for answering lowest common
ancestor (LCA) queries of~\cite{ScV-87}
and the parallel Accelerated Centroid Decomposition
of \cite{CV-86}. Since the parallelization is mostly
of technical nature it is not described in the paper.

We establish a trade-off between the time
needed to answer a query and the preprocessing time.
This trade-off is very strict. For
example, if we want each Linear Product query to be answered in at
most two steps we have to preprocess the input $\bigtheta{n\log n}$
time, but if we allow four computation steps per each query we may
preprocess the input only $\bigo{n\log^* n}$ time. On the other
hand, if we want the preprocessing to be in linear time, we must
allow $\bigtheta{\alpha (n)}$ computation steps for each query.
Similar trade-off appears in~\cite{DDPW-83} as the trade-off
between the the depth and the size of superconcentrators and
in~\cite{CFL-83a,CFL-83b} as the trade-off between the depth and the size
of certain unbounded fanin circuits.

\cite{Ta-79} considers an off-line version of the Tree Product problem.
He gives an \emph{off-line} sequential algorithm based on path compression
for \emph{commutative} semigroups
which runs in time $\bigo{(m+n) \alpha^T (m+n,n)}$, where $m$ is the number of
off-line queries, $n$ is the number of vertices and $\alpha^T (\cdot,\cdot)$ is
a function closely related to the inverse Ackermann function.
Our algorithm, which is completely different,
has several advantages relative to Tarjan's algorithm:
(1) It is on-line. (2) It works for any associative semigroup.
(3) We do not consider the amortized complexity. That is, the total
complexity is determined by the preprocessing time and the number
of queries multiplied by the maximum answering time of a query.
(4) We give an algorithm scheme which enables us to design a
preprocessing algorithm for any desired answering time.
Note that we can design an algorithm which
achieves the same time complexity as~\cite{Ta-79}. For this we
preprocess the input $\bigo{(m+n)}$ time
and then answer each of the $m$ queries in at most
$\bigo{\alpha^T (m+n,n)}$ steps.

Our algorithms use the divide-and-conquer technique.
Specifically, we decompose the size $n$ problem into subproblems of
smaller size and show that after investing linear time and space we
may consider each subproblem independently. This framework gives us,
in fact, running time which converges to $\bigo{n \alpha (n)}$.

To get some intuition of the queries we give
an example. Let the semigroup $S$
be the set of real numbers and let the
semigroup operation be the minimum operation. In this
interpretation the Linear Product problem gets the following
meaning. Suppose we are given a vector of
$n$ real numbers $a_1 ,\ldots, a_n$.
We want to find, as fast as possible, the minimum number in any
sub-vector. That is, to find $\min\lrc{a_i , a_{i+1} ,\ldots, a_{j-1} , a_j}$,
for any given $1\le i\le j\le n$.
Similarly, in the Tree Product problem,
we are given a tree with a real number associated with each of
its vertices.
We want to find, as fast as possible, the minimum number along any
path in the tree.

The defined queries have many applications. For example, consider a
communication network, where the nodes are connected using a
spanning tree. Assume that each link of the network has a specified
capacity. Each time we want to communicate from one point to
another we have to know the maximum message size we can send.
This size equals the minimum capacity along the path connecting
the two points and so it can be found by answering a suitable Tree
Product query in the given tree.

\iffalse
Another application is concerned with Database
retrieval (cf. [Da-77]).
Suppose, for example,
we have a Database of employees ordered according to the
date in which they joined the company.
We want to know, e.g.,
which one of the employees who work in the company between 10 to 15
years gets the highest salary. Obviously, this corresponds to a
suitable Linear Product query. Similar queries, which are often
called $range~queries$, can be formulated in many cases as Linear
or Tree Product queries.
\fi

\cite{Ta-79} gives three applications for his off-line algorithm.
(1) Finding maximum flow
values in a multiterminal network. (2) Verifying minimum spanning
trees. That is, given a graph and a spanning tree, verify whether
the spanning tree is minimal. (3) Updating a minimum spanning tree
after increasing the cost of one of its edges. Naturally,
we can use our algorithm for these applications also.

The rest of the paper is organized as follows.
In Section 2 we give upper and lower bounds for the
Linear Product problem.
In Section 3 we present preprocessing algorithms for the
Tree Product problem.
Section 4 contains some concluding remarks and open problems.

\section{The Linear Product Query}

Suppose we are given a semigroup $(S,\myotimes)$.
Let $s_1 ,\ldots, s_n$ be elements of $S$.
In this section we examine the amount of sequential
preprocessing needed for answering
queries of the form, ``What is the product
$s_i \myotimes s_{i+1} \myotimes \cdots \myotimes s_{j-1} \myotimes s_j$?'' for any given
$1\le i\le j\le n$.

\subsection{The upper bound}

At first we describe preprocessing algorithms for answering
Linear Product queries in at most one and two steps.
Afterwards, we describe a preprocessing algorithm for any $k>2$ steps.
This algorithm uses the algorithm for $k-2$ steps as a subroutine.

\par\noindent\emph{The preprocessing algorithm for one step.}
Observe that the best preprocessing algorithm for one step is the
naive algorithm which precomputes all the necessary products in
advance. Clearly, this takes $\bigo{n^2}$ time and space.

\par\noindent\emph{The preprocessing algorithm for two steps.}
Let $\ell$ be $\floor{n/2}$. We precompute all the products
$s_i \myotimes \cdots \myotimes s_\ell$, for $1\le i< \ell$, and
$s_{\ell+1} \myotimes \cdots \myotimes s_j$, for $\ell+1< j\le n$.
This can be done in linear time and space.
Suppose we are given the query
$s_i \myotimes s_{i+1} \myotimes \cdots \myotimes s_{j-1} \myotimes s_j$. If $j=\ell$ or $i=\ell+1$
then we can answer the query in one step using the precomputed products.
If $i\le \ell$ and $j>\ell$
then we can answer the query in two steps by
computing the product of the
precomputed $s_i \myotimes \cdots \myotimes s_\ell$ and $s_{\ell+1} \myotimes \cdots \myotimes s_j$.
Thus, the rest of the preprocessing should be aimed for answering
queries of the form
$s_i \myotimes s_{i+1} \myotimes \cdots \myotimes s_{j-1} \myotimes s_j$, where either
$1\le i\le j<\ell$, or $\ell+1<i\le j\le n$.
This is done recursively using the same method.
The total preprocessing time and space is given by the
recurrence: $T_2 (n)\le 2 T_2 (n/2)+n$, whose solution is
$n\log n$. Hence, the preprocessing takes $\bigo{n \log n}$ time and
space.

\noindent
Notice that when we answer a query we have to be able to retrieve
each of the precomputed products whose multiplication gives the result
with no overhead. To simplify the retrieval we modify the
preprocessing algorithm (without changing its complexity bounds).
We start the algorithm with $\ell = 2^{\floor{\log n}}$
instead of $\ell = \floor{n/2}$ and continue in the same manner when
we recur. One can easily see that after this modification we can
retrieve the precomputed products required for answering a query
$s_i \myotimes s_{i+1} \myotimes \cdots \myotimes s_{j-1} \myotimes s_j$ by performing $\log n$-bit
operations on the indices $i$ and $j$. (In case these operations are
not part of our machine's repertoire, look-up tables for each
missing operation are prepared in linear time and linear space as
part of the preprocessing. These tables will be used to perform the
missing operations in constant time.)

To describe our algorithm for $k>2$ steps, we shall need to
define some very rapidly growing and very slowly growing functions.
Following~\cite{Ta-75}, we define:
\begin{align*}
  A(0,j) &=2j & \mbox{for } & j\ge 1 \\
  A(i,0) &=1  & \mbox{for } & i\ge 1 \\
  A(i,j) &= A(i-1,A(i,j-1)) & \mbox{for } & i,j\ge 1
\end{align*}

Similarly, we define:
\begin{align*}
  B(0,j) &=j^2 & \mbox{for } & j\ge 1 \\
  B(i,0) &=2   & \mbox{for } & i\ge 1 \\
  B(i,j) &= B(i-1,B(i,j-1))  & \mbox{for } & i,j\ge 1
\end{align*}

\noindent
For $i\ge 0$, define $\lambda (2i,x) = \min\lrc{j\,|\, A(i,j)\ge x}$
and $\lambda (2i+1,x) = \min\lrc{j\,|\, B(i,j)\ge x}$.

\noindent
We give the first five functions explicitly.
$\lambda (0,x) =\ceil{x/2}$,
$\lambda (1,x) =\ceil{\sqrt{x}}$, $\lambda (2,x) =\log x$,
$\lambda (3,x) =\log \log x$, $\lambda (4,x) =\log^* x$.
Observe that
$\lambda (i,x) =\min\lrc{j\,|\, \lambda^{(j)} (i-2,x)\le 1}$,
for $i\ge 2$,
where $\lambda^{(1)} (i,x) =\lambda (i,x)$ and
$\lambda^{(j)} (i,x) =\lambda  (i, \lambda^{(j-1)} (i,x))$.

\noindent\textit{Remark:}
The bound of $\bigtheta{n \lambda (k,n)}$
on the time and space of the preprocessing algorithm for $k$
steps is valid for $k >1$.
Notice that in order to answer Linear Product queries
in \emph{one} step we have to invest $\bigtheta{ n^2 }$ time and
space and not $\bigtheta{ n \lambda (1,n) } = \bigtheta{n^{1.5}}$.

Finally, we define the inverse Ackermann function
$\alpha (x) =\min\lrc{j\,|\, A(j,j)\ge x}$.

\noindent\textit{The preprocessing algorithm for $k>2$ steps.}
The preprocessing algorithm for $k>2$ steps
runs in $\bigo{n \lambda (k,n)}$ time and space.
Let $\ell$ be $\lambda (k-2,n)$ and let $m$ be $n/\ell$.
(To simplify the presentation we assume that $m$ is an integer.)
For each $x=1,\ldots,m$ we precompute all the products
$s_i \myotimes \cdots \myotimes s_{\ell x}$, for $\ell(x-1)<i<\ell x$.
For each $x=1,\ldots,m-1$ we precompute all the products
$s_{\ell x+1} \myotimes \cdots \myotimes s_j$, for $\ell x+1<j<\ell(x+1)$.
This can be done in time and space which are proportional to $2n$.
Let $\bar{s}_x$ be $s_{\ell(x-1)+1} \myotimes \cdots \myotimes s_{\ell x}$, for $x=1,\ldots,m$.
We preprocess the $m$ elements $\bar{s}_1 ,\ldots, \bar{s}_m$, using the
preprocessing algorithm for $k-2$ steps.
This is done in time and space which are proportional to
$(k-2)m \lambda (k-2,m))\le (k-2)n$. (Note that this holds also for
$k=3$.) Suppose we are given the query
$s_i \myotimes s_{i+1}  \myotimes \cdots \myotimes s_{j-1} \myotimes s_j$ (for $i<j$).
Let $x$ be $\ceil{i/\ell}$ and let $y$ be $\floor{j/\ell}$.
If $x=y$ then we can compute the product in at most
two steps as in the algorithm for two steps.
If $x<y$ then we answer the query as follows:
\par\noindent
(i) We compute $\bar{s}_{x+1} \myotimes \cdots \myotimes \bar{s}_y$
in $k-2$ steps. (ii) We multiply
the products: $s_i \myotimes \cdots \myotimes s_{\ell x}$,
$\bar{s}_{x+1} \myotimes \cdots \myotimes \bar{s}_y$ and $s_{\ell y+1} \myotimes \cdots \myotimes s_j$.
Thus, the rest of the preprocessing should be aimed for answering
queries of the form
$s_i \myotimes s_{i+1}  \myotimes \cdots \myotimes s_{j-1} \myotimes s_j$, where
$\ell(x-1)<i<j<\ell x$, for some $1\le x\le m$.
This is done recursively using the same method.
The total preprocessing time and space is given by the
recurrence:
$T_k (n)\le \frac{n}{\lambda (k-2,n)}T_k ( \lambda (k-2,n))+kn$.
It is not difficult to verify that the solution of this recurrence
is $kn \lambda (k,n)$. Since $k$ is constant we have that the
total preprocessing time and space is $\bigo{n \lambda (k,n)}$.

We show how to retrieve the precomputed products required for
answering a query
%$s_i \myotimes s_{i+1}  \myotimes \cdots \myotimes s_{j-1} \myotimes s_j$.
$s_i \myotimes s_{i+1}  \myotimes \cdots \myotimes s_j$.
We start by retrieving the first and last required products.
Again, let $\ell = \lambda (k-2,n)$ and $m=n/\ell$.
We distinguish between two cases.
\textit{(Case A)} $x=\ceil{i/\ell} \le y=\floor{j/\ell}$. In this case the first
product is $s_i \myotimes \cdots \myotimes s_{\ell x}$ and the last is
$s_{\ell y+1} \myotimes \cdots \myotimes s_j$.
\textit{(Case B)} Not Case A. That is, $\ell(x-1)<i<j<\ell x$,
for some $1\le x\le m$. Or, in words, $i$ and $j$ belong to the
same block of size $\ell$.
In this case we retrieve the required products
using a look-up table. Observe that these products are defined
(within their block) by $i\bmod \ell$ and
$j\bmod \ell$. Hence, the needed look-up table is
of size $\ell \times \ell$. This table can be
computed in linear time and space during the preprocessing stage.
The rest of the required products are retrieved recursively in the
same method. Notice that we have $\ceil{k/2}$ recursion levels.
Hence, the retrieval is done with no time or space overhead.

We conclude the description of the upper bounds
by presenting the
best linear preprocessing algorithm. This linear time and space
preprocessing algorithm enables us to answer Linear Product
queries in $\bigo{\alpha (n)}$ steps.

We start by describing a simple linear preprocessing algorithm
(which is not the best). We use a balanced binary tree
whose leaves are $s_1 ,\ldots, s_n$. For each internal vertex we compute
the product of its descendent leaves. Clearly, this can be done in
linear time and space. It can be easily verified that using these
precomputed products we can compute
$s_i \myotimes s_{i+1}  \myotimes \cdots \myotimes s_{j-1} \myotimes s_j$ for any given
$1\le i\le j\le n$ in at most $2 \ceil{\log n}$ steps.
This is done, simply, by ``climbing'' from the leaf (which
represents) $s_i$ and the leaf (which represents) $s_j$ to the
lowest common ancestor (LCA) of these leaves.

We combine this simple preprocessing algorithm with the
preprocessing algorithm described above to get the best linear
preprocessing algorithm. Again, we use the divide-and-conquer
technique. Let $\ell=2 \alpha^2 (n)$.
We partition $s_1 ,\ldots, s_n$ into $m=n/\ell$ blocks of
size $\ell$ each.
We preprocess each block using the above simple algorithm.
This takes linear time and space.
It enables us to answer intra-block queries in at most
$2 \ceil{\log \ell} = \bigo{\log \alpha (n)}$ steps.
We aim to preprocess the input so
that we will be able to answer inter-block queries in
$\bigo{\alpha (n)}$ steps. For this, we define $\bar{s}_1 ,\ldots, \bar{s}_m$ as before
(that is, $\bar{s}_x=s_{\ell(x-1)+1} \myotimes \cdots \myotimes s_{\ell x}$)
and preprocess these $m$ elements using
the preprocessing algorithm for $2 \alpha (n)$
steps described above. In addition we precompute
for each $x=1,\ldots,m$ all the products
$s_i \myotimes \cdots \myotimes s_{\ell x}$, for $\ell(x-1)<i<\ell x$, and
for each $x=1,\ldots,m-1$ all the products
$s_{\ell x+1} \myotimes \cdots \myotimes s_j$, for $\ell x+1<j<\ell(x+1)$.
This enables us to answer inter-block queries in at most
$2 \alpha (n) +2$ steps.
Observe that the preprocessing takes
$2 \alpha (n)m \lambda (2 \alpha (n),m)=
2 \alpha (n) \cdot n/\lrp{2 \alpha^2 (n)}\cdot \alpha (n)=n$ time and space.
(\emph{Remark:} The preprocessing algorithm for $k$
steps requires $kn \lambda (k,n)$ time and space.
Since in our case $k=2 \alpha (n)+2$ is not a
constant we had to take it into consideration.)
This implies that we can answer queries in at most $\bigo{\alpha (n)}$
steps after linear preprocessing. Note that this is best possible
since, by our lower bound proven below,
$\bigomega{\alpha (n)}$ steps is the best number of steps
achievable even in case
we allow a preprocessing of $n \alpha (n)$ time and space.

\subsection{The lower bound}

We show that a preprocessing of $\bigomega{n \lambda (k,n)}$ time and
space is needed for answering Linear Product queries in at most
$k$ steps. We have the following assumptions.
(1) Given two elements $a$ and $b$ in the semigroup $S$
we can compute $a \myotimes b$ in constant time.
(2) The only available operation on the semigroup elements
is the semigroup operation.
(\emph{Remark:}
Assumption (2) is crucial for the lower bound. To see this, note that
if the given semigroup is a group we can perform $n$ prefix
computations on the input in linear time and then answer
Linear Product queries in constant time using the inverse operation
of the group.)

For proving the lower bound we prove a stronger result, which is of
independent interest. For two integers $i\le j$,
we denote the set of integers $\lrc{i,i+1,\ldots,j-1,j}$ by $[i,j]$.
(We shall refer to it as the \emph{integer interval} $[i,j]$.)
A set $\calP$ of subsets of $[1,n]$ is said to be a \emph{$k$-covering} set of
$[1,n]$ if each integer interval contained in $[1,n]$
(i.e., each integer interval $[i,j]$, for $1\le i\le j\le n$)
is the union of at most $k$ subsets in $\calP$.
We want to find a lower bound for the minimum possible cardinality of
a $k$-covering set of $[1,n]$, denoted $P_k (n)$.
We prove that
$P_k (n)=\bigomega{n \lambda (k,n)}$, for $k\ge 2$.
(Note that the size of any $1$-covering set is $\bigomega{n^2}$.)
We claim that $P_k (n)$ is a lower bound on the preprocessing needed
for answering Linear Product queries in at most $k$ steps.
To prove this claim associate each product precomputed in the
preprocessing algorithm with the
set of indices of the elements consisting it. Notice that this
gives a $k$
covering set as each integer interval $[i,j]$ must be the union of
at most $k$ of these sets (corresponding to the precomputed
products whose product is $s_i \myotimes s_{i+1}  \myotimes \cdots \myotimes s_{j-1} \myotimes s_j$).
Moreover,
our lower bound is stronger than the lower bound for the Linear
Product problem in two aspects:
(i) We consider coverings by \emph{subsets}, while for the Linear
Product problem we may consider only coverings by integer
\emph{intervals}.
(ii) We consider any coverings, while for the Linear
Product problem we may consider only \emph{exact} coverings. That is,
only coverings for which each integer subinterval is the union of
at most $k$ \emph{pairwise~disjoint} subsets.
This stronger result, implies that we can not improve our
algorithm even if the semigroup $S$ is commutative and/or
consists only of idempotent elements.
(That is, $a \circ a=a$, for every element $a\in S$.)
An example of a semigroup which is both commutative and
consists only of idempotent elements
is any set of numbers with
the operation maximum or minimum.

The lower bound is proven inductively. We start by proving the lower
bound for $k=2$ and then prove it for any $k>2$.

\par\noindent\textit{The lower bound for $k=2$.}
We show that $P_2 (n)$
satisfies the recurrence
$P_2 (n)\ge P_2 (\ceil{n/2} -1)+P2 (\floor{n/2})+\floor{n/2}$.
The lower bound $\bigomega{n\log n}$ follows.
Let $\ell$ be $\floor{n/2} +1$. Partition $[1,n]$ into two subintervals
$I_1=[1,\ell-1]$ and $I_2=[\ell+1,n]$.
Clearly, a $2$-covering set of $[1,n]$ must contain $2$-covering
sets of $I_1$ and $I_2$. Moreover, these $2$-covering sets are
disjoint as the $2$-covering set
of $I_1$ (resp. $I_2$) consists only of subsets of $I_1$ (resp. $I_2$).
We show that any $2$-covering set of $[1,n]$
must contain $\floor{n/2}$ additional subsets.
Each one of these additional subsets contains either the element $\ell$
or elements from $I_1$ and from $I_2$.
Let $\calP$ be a $2$-covering set of $[1,n]$.
We distinguish between two cases:
\par\noindent\textit{(Case A)}
Each element $x$ in $I_1$ belongs to a subset $Q\in \calP$
such that: (i) $x$ is the minimal element in $Q$. (ii) $Q$ contains
elements which are $\ge \ell$. Clearly, in this case we have
$\floor{n/2}$ different subsets containing elements both from $I_1$
and from $I_2 \cup \lrc{\ell}$.
\par\noindent\textit{(Case B)}
Not Case A. That is, there exists an element $x$ in $I_1$
such that each subset $Q\in \calP$ which contains $x$ as its
minimal element does not contain elements which are $\ge \ell$.
Consider the intervals $[x,i]$, for
$i\ge\ell$. Each such interval must be the union of at most
two subsets in $\calP$. Since no subset $Q\in \calP$ satisfies
conditions (i) and (ii) of Case A,
each interval $[x,i]$ must be the union of
exactly two subsets in $\calP$.
One subset, say $Q_1$, must contain $x$ as its
minimal element, and the other subset, say $Q_2$, must contain $\ell$.
Also, note that the maximum element in $Q_2$ must be $i$. Hence,
we have $\ceil{n/2}$ different subsets containing $\ell$.

\par\noindent\textit{The lower bound for $k>2$.}
We show that $P_k (n)$ satisfies the recurrence
\[
P_k (n)\le \frac{n}{\lambda (k-2,n)}\cdot P_k (\lambda (k-2,n)-1)+\bigomega{n}.
\]

The lower bound $\bigomega{n\lambda (k,n)}$ follows.
Let $\ell$ be $\lambda (k-2,n)$ and let $m$ be $n/\ell$. (Again, we assume
that $m$ is an integer.) Partition $[1,n]$ into $m$ subintervals:
$I_j=[\ell(j-1)+1,\ell j-1]$, for $j=1, \ldots, m$.
Clearly, a $k$-covering set of $[1,n]$ must contain $k$-covering
sets of each $I_j$.
Moreover, these $k$-covering sets are
disjoint as the $k$-covering set
of $I_j$ consists only of subsets of $I_j$.
We show that any $k$-covering set of $[1,n]$
must contain $\bigomega{n}$ additional subsets.

Let $\calP$ be a $k$-covering set of $[1,n]$.
A subset $Q\in\calP$ is \emph{global} if it is not contained in any
subinterval. An element $x\in I_j$ is \emph{global}
if it is an extremal (minimal or maximal) element in some global
subset.  Finally,
a subinterval is \emph{global} if all of its elements are global.
We show that $\calP$ contains $\bigomega{n}$ global subsets.
We distinguish between two cases:
\par\noindent\textit{(Case A)}
There are $\floor{m/2}$ global subintervals.
Each global subinterval contains $\ell-1$ global elements.
Note that each global element corresponds to at least one global
subset, and that each global subset may correspond to at most two
global elements. Hence, there must
be at least $\frac 14 m (\ell-1)=\bigomega{n}$ global subsets.
\par\noindent\textit{(Case B)}
Not Case A. That is, there are $\floor{m/2}$ subintervals
which are not global. Note that each nonglobal subinterval
contains at least one nonglobal element. That is, an element which
is not extremal in any global subset. Let
$x\in I_{j_1}$ and $y\in I_{j_2}$ be nonglobal elements, where $j_1<j_2$.
Consider the interval $[x,y]$.
It must be the union of at most
$k$ subsets of $\calP$. Since no global subset contains $x$ (resp.
$y$) as its minimal (resp. maximal) element, at least one of these
subsets must be contained in $I_{j_1}$ (resp. $I_{j_2}$). Hence
the union of the rest of the subsets contains the interval
$[\ell j_1 , \ell( j_2 -1)]$. This implies that if we omit
from the subsets in $\calP$
all the elements which are not of the
form $\ell x$, for some $x=1,\ldots,m-1$ such that $I_x$ is nonglobal,
we are left with a
$(k-2)$-covering set for the set $\lrc{\ell x\,|\,I_x \mbox{ is nonglobal}}$.
By the inductive hypothesis this set must contain
$\bigomega{(m/2)\lambda (k-2,m/2)}=\bigomega{n}$ nonempty subsets.
Note that each such subset corresponds to at least one global
subset in $\calP$. Hence, $\calP$ contains $\bigomega{n}$ global subsets.

\section{The Tree Product Query}

Let $T$ be an unrooted tree with an element of $S$ associated with
each of its vertices. We want to answer on-line
queries of the form, ``What is the product
of the elements associated with the vertices along the path from $u$ to
$v$?'' for any pair of vertices $u$ and $v$ in $T$.
(We denote such a query $\TreeProduct(u,v)$.)
In this section we present an $\bigo{n \lambda (k,n)}$ time and space
preprocessing algorithm for answering Tree Product queries in at
most $2k$ steps, where $n$ is the number of vertices and $k\ge 2$ is a
fixed parameter. We also show a linear time and linear space
preprocessing algorithm for answering Tree Product queries in
$\bigo{  \alpha (n)}$ steps.

We start by showing that it is sufficient to preprocess $T$
only in order to answer queries of the form $\TreeProduct(u,v)$,
where either $u$ is the ancestor of $v$
or vice versa, for an arbitrarily chosen root $r$ of $T$.
Suppose we are given a query
$\TreeProduct(u,v)$ such that neither $u$ is an ancestor of $v$
nor $v$ is an ancestor of $u$. We answer it in three stages.
(1) We find the lowest common ancestor of $u$ and $v$ (denoted
$LCA(u,v)$).
(2) We compute $\TreeProduct(u,LCA(u,v))$.
(3) We compute $\TreeProduct(w,v)$, where $w$ is the child of $LCA(u,v)$ that is also an ancestor of $v$.
Clearly,
\[
\TreeProduct(u,v)=\TreeProduct(u,LCA(u,v)) \myotimes \TreeProduct(w,v)).
\]
For computing $LCA(u,v)$ we preprocess $T$ using the linear time and
space preprocessing algorithm of \cite{HT-84} or the simplified
preprocessing algorithm of \cite{ScV-87}. These preprocessing algorithms
enable us to answer queries of the form,
``Which vertex is the lowest common ancestor (LCA) of
$u$ and $v$?'' for any pair of $u$ and $v$ in $T$, in constant time.
Below, we
present a preprocessing algorithm for answering queries of the
form $\TreeProduct(u,v)$, where $u$ is the ancestor of $v$ in at
most $k$ steps. Our preprocessing algorithm
takes $\bigo{n \lambda (k,n)}$ time and space.
The preprocessing algorithm for
answering queries of the form $\TreeProduct(u,v)$, where
$v$ is the ancestor of $u$ is similar. (Note that when the
semigroup is not commutative it is possible that
$\TreeProduct(u,v)\ne \TreeProduct(v,u)$.)
Combining both algorithms results in an $\bigo{n \lambda (k,n)}$
time and space preprocessing
algorithm for answering a general Tree Product query in at most
$2k$ steps.

\subsection{High-level description of the preprocessing algorithm}

Our preprocessing algorithm uses a
divide-and-conquer technique as the preprocessing algorithm for
answering Linear Product queries. That is, we partition the tree
into $\bigo{n/ \lambda (k-2,n)}$
connected components of size $\bigo{\lambda (k-2,n)}$ each
and show that after investing linear time and space work we may
consider each connected component independently.
We decompose the size $n$ problem into problems of size
$\bigo{\lambda (k-2,n)}$ each in four stages.

\par\noindent\textit{Stage 1.}
We binarize the tree $T$ using
a well-known transformation as follows.
For each vertex $v$ in $T$ of outdegree $d>2$, where $w_1 ,\ldots, w_d$
are the children of $v$, we replace $v$ with the new vertices
$v_1 ,\ldots, v_{d-1}$. We make $v_i$ the parent of $v_{i+1}$ and $w_i$,
for $i=1,\ldots,d-2$ and $v_{d-1}$ the parent of $w_{d-1}$ and $w_d$.
Let $B$ be the resulting rooted binary tree.
Note that the number of vertices in $B$ is at most twice the number
of vertices in $T$.
Finally, we associate the unit element of $S$ with each new vertex.
(Note that in case $S$ lacks a unit element we can simply
add such an element to $S$.)
Clearly, this can be done in linear time and space.

\par\noindent\textit{Stage 2.}
We partition $B$ into $\bigo{n/ \lambda (k-2,n)}$
connected components of size $\bigo{\lambda (k-2,n)}$ each by removing
$\bigo{n/ \lambda (k-2,n)}$ edges.
(For $k=2$ we partition $B$ into exactly two components of
size between $n/3$ and $2n/3$ each.)
The existence of such a partitioning is guaranteed by the separator
theorem of~\cite{LT-79} for the family of trees with a maximum degree three.
The partitioning can be done in linear time and space
using Depth First Search as shown in~\cite{Fr-85}.

Let $C$ be one of the resulting connected components of $B$.
Note that $C$ is also a rooted binary tree.

\par\noindent\textit{Stage 3.1.}
For each vertex $x$ in $C$ we compute
$\TreeProduct( r_C ,x)$, where $r_C$ is the root of $C$.
This can be done in linear time and space
using, e.g., Breadth First Search.

\par\noindent\textit{Stage 3.2.}
For each vertex $x$ in $C$ such that at least one child of $x$
(in $B$) does not belong to $C$ and for each ancestor $y$ of $x$ in
$C$, we compute $\TreeProduct(y,x)$.
The products are computed in constant time per product
by ``climbing'' from each such vertex $x$
to the root of its component $r_C$.
Note that the total number of such vertices $x$ is
$\bigo{n/ \lambda (k-2,n)}$, also, the number of ancestors of each
such vertex in its component is $\bigo{\lambda (k-2,n)}$.
Hence, the total number of products computed in this stage is
$\bigo{n}$. Thus, this stage takes also linear time and space.

Let $v$ be a vertex in $T$. Denote by
$C(v)$ the connected component which contains $v$ and
by $F_B (v)$ the parent of $v$ in $B$.
We define a new rooted tree $\bar{B}$ as follows.
The vertices of $\bar{B}$ correspond to the connected
components of $B$. The root of $\bar{B}$ is $C(r)$, where $r$ is the
root of $B$. For every edge $(v, F_B (v))$ in $B$ such that
$C(v)\ne C( F_B (v))$ we make $C( F_B (v))$ the parent of $C(v)$.
We associate an element from $S$ with each vertex of $\bar{B}$ as follows.
(i) The unit element of $S$ is associated with the root of $\bar{B}$.
(ii) To each other vertex $C$ in $\bar{B}$ we associate
$\TreeProduct( r_D , F_B ( r_C ))$,
where $r_D$ is the root of the connected component $D=F_{\bar{B}} (C)$.
(Notice that these products were computed in the previous stage.)

\par\noindent\textit{Stage 4.}
We perform the preprocessing algorithm for answering Tree
Product queries in at most $k-2$ steps on $\bar{B}$.
(This is done only if $k>2$.)
This takes $\bigo{\lambda (k-2,n) (n/ \lambda (k-2,n))} =\bigo{n}$
time and space.

\par\noindent\textit{The validity of the decomposition algorithm.}
We show how to answer a query $\TreeProduct(u,v)$ in $T$, such
that $u$ and $v$ belong to different components in at most $k$
steps. Recall that $u$ is an ancestor of $v$.
If $C(u)$ is the parent of $C(v)$ then
$\TreeProduct(u,v)$ is the product of the precomputed
$\TreeProduct(u, F_B ( r_{C(v)} ))$ and $\TreeProduct ( r_{C(v)} ,v)$.
Suppose $k>2$.
Let $x$ be the last vertex of $C(u)$ which appears along
the path from $u$ to $v$ in $B$
and let $C$ be the \emph{grandchild} of $C(u)$ which appears along
the path from $C(u)$ to $C(v)$ in $\bar{B}$.
$\TreeProduct(u,v)$ is the product of
(i) $\TreeProduct(u,x)$ in $B$, precomputed in Stage 3.2
(ii) $\TreeProduct(C,C(v))$ in $\bar{B}$. Using the preprocessing of
Stage 4, this product can be computed in at most $k-2$ steps.
(iii) $\TreeProduct ( r_{C(v)}  ,v)$ in $B$, precomputed in Stage 3.1.
In order to be able to retrieve these precomputed products
within the stated complexity bounds
we must be able to find $x$ and $C$ in constant time.
Using the ideas of the algorithms of~\cite{HT-84} and~\cite{ScV-87}
we can preprocess $\bar{B}$ in linear time and space
such that we would be able to find
the ancestor of $w$ whose distance from the root is $d$
in constant time, for any vertex $w$ in $\bar{B}$ and any distance $d$.
This enables us to find $C$ whose distance from $C(v)$ is given in
constant time. Given $C$, we can also find
$x=F_B ( r_D )$,
where $r_D$ is the root of the connected component $D=F_{\bar{B}} (C)$,
in constant time.

All the above discussion applies to $k\ge 2$, i.e., to answering
time of $2k\ge 4$ per query.
It is worth noting that using similar ideas we can design
an $\bigo{n \log n}$ time and space
preprocessing algorithm for answering Tree Product queries in at
most two steps per query and an $\bigo{n \log\log n}$ time and space
preprocessing algorithm for answering Tree Product queries in at
most three steps per query.

We conclude this section by describing a linear time and
space preprocessing algorithm for answering Tree Product queries
in $\bigo{\alpha (n)}$ steps.

As in the description of the best linear preprocessing algorithm
for answering Linear Product queries,
we start by describing a preliminary linear preprocessing algorithm
which is not the best.

\par\noindent\textit{The preliminary linear preprocessing algorithm.}
Suppose we are given a tree $T$, rooted at $r$.
We present a linear time and space preprocessing algorithm which
enables us to answer queries of the form $\TreeProduct(u,v)$, for
any pair of vertices $u$ and $v$, such that $u$ is an ancestor of
$v$, in $\bigo{\log n}$ steps. Following~\cite{AHU-76,Ta-75,HT-84},
we partition $T$ into a collection of
disjoint paths, as follows. For each vertex $v$ in $T$, let
$SIZE(v)$ be the number of its descendants (including itself).
Define an edge $(v,u)$ (where $u$ is the parent of $v$) to be
\emph{heavy} if $2SIZE(v)\ge SIZE(u)$ and \emph{light} otherwise. Since the
size of a vertex is one greater than the sum of the sizes of its
children, at most \emph{one} heavy edge exits from each vertex. Thus, the
heavy edges partition the vertices of $T$ into a collection of
\emph{heavy~paths}. (A vertex with no entering or exiting heavy edge is
a single-vertex heavy path.)
Define the \emph{head} of a heavy path to be the vertex
which is closest to $r$ in this heavy path.

Let $u$ and $v$ be two vertices in
$T$, such that $u$ is an ancestor of $v$. One can easily verify
the following two facts. (1) The vertices along the (unique)
path in $T$ between $u$ and
$v$ are partitioned by the heavy edges into at most $\ceil{\log n}$
heavy sub-paths.
(2) Each such heavy sub-path, except possibly the first, starts with
the head of its corresponding heavy path.

We are ready now to describe the preliminary preprocessing algorithm.
It has three stages each taking linear time and linear space.
(1) We partition the input tree $T$ into heavy paths as described
above. (2) We preprocess each heavy path
using the simple linear preprocessing algorithm for answering Linear
Product queries described in Section 2.
(3) For each head $u$ of a heavy path and
for each vertex $v$ in its heavy
path we compute $\TreeProduct(u,v)$.

Suppose we are given a query $\TreeProduct(u,v)$ we show how to
answer it in $\bigo{\log n}$ steps. We have two possibilities.

\par\noindent\textit{(Possibility A)}
$u$ and $v$ are in the same heavy path. In this
case we can answer the query using the preprocessing of each heavy
path done in Stage 2.
\par\noindent\textit{(Possibility B)}
$u$ and $v$ are not in the same heavy path.
Recall that the path from $u$ to $v$ is partitioned into
at most $\ceil{\log n}$ heavy sub-paths and that
each such heavy sub-path, except possibly the first, starts with
the head of its corresponding heavy path. Thus, to
compute $\TreeProduct(u,v)$ we multiply (i) the $\bigo{\log n}$
precomputed products, computed in Stage 2, which give the
product of the vertices along the first sub-path. (ii) the
$\bigo{\log n}$
precomputed products, computed in Stage 3, which give the
product of the vertices along the rest of the sub-paths.
(One precomputed product per each such sub-path.)

The linear preprocessing algorithm for answering Tree Product
queries has five stages.
(1) We binarize the input tree $T$.
Let $B$ be the resulting binary tree. (2) We decompose $B$
into connected components of size $\bigo{\alpha^2 (n)}$ each.
As in the decomposition algorithm, described in the start of this
section,
we compute the following for each connected component $C$ of $B$.
(3.1) For each vertex $x$ in $C$ we compute
$\TreeProduct( r_C ,x)$, where $r_C$ is the root of $C$.
(3.2) For each vertex $x$ in $C$ such that at least one child  of $x$
(in $B$) does not belong to $C$ and for each ancestor $y$ of $x$ in
$C$, we compute $\TreeProduct(y,x)$.
We define the tree $\bar{B}$ as in the decomposition algorithm above.
Note that $\bar{B}$ has $\bigo{n/\alpha^2 (n)}$ vertices.
(4) We preprocess $\bar{B}$ in $\bigo{n}$ time
and space using our preprocessing algorithm for answering Tree
Product queries in at most $\bigo{\alpha (n)}$ steps.
As shown before,
this preprocessing together with the computation done in Stage 3
enable us to answer
inter-component queries in at most $\bigo{\alpha (n)}$ steps.
(4) We preprocess each component in linear time and linear space using
the preliminary linear preprocessing algorithm.
This preprocessing enables us to answer
intra-component queries in $\bigo{\log\alpha (n)}$ steps.
Thus, we can answer any query in $\bigo{\alpha (n)}$ steps.

\section{Open Problems}

We presented efficient preprocessing algorithms for
answering product queries. Under reasonable assumptions
these algorithms are optimal. Note that they apply only to
\emph{static} input.
The most important open problem is what can be done when the input
is not static. In the Linear Product case,
we see three kinds of dynamic operations:
(i) changing the value of an element. (ii) adding a new element
(iii) deleting an element. The simple linear preprocessing
algorithm, described in Section 2, can be easily adapted
to the dynamic case. It gives a
linear preprocessing algorithm which enables us to perform each of
the above three dynamic operations and also to answer Linear
Product queries in $\bigo{\log n}$ time. We do not know whether this is
best possible and we are also unable to prove any nontrivial
lower bound or trade off between preprocessing time and processing
time. For the Tree Product case, the possible dynamic operations are:
(i) changing the value of an element (ii) linking two trees by
adding a new edge. (iii) cutting a tree by deleting an edge.
Using the ideas of~\cite{HT-84} and~\cite{ST-83} we can design a linear
preprocessing algorithm for this case which will also enable us to
perform each dynamic operation and to answer Tree
Product queries in $\bigo{\log n}$ time. Again,
we do not know whether this is best possible.

Another direction for future work is to find more applications
where the described preprocessing algorithms for answering product
queries can be used.

\section*{Acknowledgements} We are grateful to Zvi Galil, Yael Maon, and
Uzi Vishkin for stimulating discussions and helpful comments.

\bibliographystyle{alpha}
\bibliography{preproc}
\end{document}